\documentclass[sigconf]{acmart}

\AtBeginDocument{%
  \providecommand\BibTeX{{%
    \normalfont B\kern-0.5em{\scshape i\kern-0.25em b}\kern-0.8em\TeX}}}

\copyrightyear{2022}
\acmYear{2022}
\setcopyright{acmcopyright}
\acmConference[WSDM '22]{Proceedings of the Fifteenth ACM International Conference on Web Search and Data Mining}{February 21--25, 2022}{Tempe, AZ, USA}
\acmBooktitle{Proceedings of the Fifteenth ACM International Conference on Web Search and Data Mining (WSDM '22), February 21--25, 2022, Tempe, AZ, USA}
\acmPrice{15.00}
\acmDOI{10.1145/3488560.3498371}
\acmISBN{978-1-4503-9132-0/22/02}

\usepackage{graphicx}
\usepackage{amsmath}
\usepackage{multirow}
\usepackage{booktabs}
\usepackage{array}
\usepackage{subfigure}
\usepackage{amssymb}
\usepackage{algorithm}
\usepackage{verbatim}
\usepackage{bm}     
\usepackage{enumitem}       

\usepackage{algpseudocode}

\begin{document}

\fancyhead{}

\title{Long Short-Term Temporal Meta-learning in Online Recommendation}


\author{Ruobing Xie}
\authornote{Both authors contributed equally to this research. Ruobing Xie is the corresponding author (ruobingxie@tencent.com).}
\affiliation{\institution{WeChat, Tencent}
\city{Beijing}
\country{China}}
\email{ruobingxie@tencent.com}

\author{Yalong Wang}
\authornotemark[1]
\affiliation{\institution{WeChat, Tencent}
\city{Beijing}
\country{China}}
\email{vinceywang@tencent.com}

\author{Rui Wang}
\affiliation{\institution{WeChat, Tencent}
\city{Beijing}
\country{China}}
\email{rysanwang@tencent.com}

\author{Yuanfu Lu}
\affiliation{\institution{WeChat, Tencent}
\city{Beijing}
\country{China}}
\email{lucasyflu@tencent.com}

\author{Yuanhang Zou}
\affiliation{\institution{WeChat, Tencent}
\city{Beijing}
\country{China}}
\email{yuanhangzou@tencent.com}

\author{Feng Xia}
\affiliation{\institution{WeChat, Tencent}
\city{Beijing}
\country{China}}
\email{xiafengxia@tencent.com}

\author{Leyu Lin}
\affiliation{\institution{WeChat, Tencent}
\city{Beijing}
\country{China}}
\email{goshawklin@tencent.com}


\begin{abstract}
An effective online recommendation system should jointly capture users' long-term and short-term preferences in both users' internal behaviors (from the target recommendation task) and external behaviors (from other tasks). However, it is extremely challenging to conduct fast adaptations to real-time new trends while making full use of all historical behaviors in large-scale systems, due to the real-world limitations in real-time training efficiency and external behavior acquisition. To address these practical challenges, we propose a novel Long Short-Term Temporal Meta-learning framework (LSTTM) for online recommendation. It arranges user multi-source behaviors in a global long-term graph and an internal short-term graph, and conducts different GAT-based aggregators and training strategies to learn user short-term and long-term preferences separately. To timely capture users' real-time interests, we propose a temporal meta-learning method based on MAML under an asynchronous optimization strategy for fast adaptation, which regards recommendations at different time periods as different tasks. In experiments, LSTTM achieves significant improvements on both offline and online evaluations. It has been deployed on a widely-used online recommendation system named WeChat Top Stories, affecting millions of users.
\end{abstract}

\ccsdesc[500]{Information systems~Recommender systems}

\keywords{recommendation, temporal meta-learning, online recommendation}


\maketitle

\section{Introduction}

Real-world industry-level recommendation systems usually need to interact with complicated practical scenarios. Large-scale online recommendation systems in super platforms such as Google and Amazon usually have the following two complexities:
(1) \textbf{\emph{user behaviors are multi-source}}. Super platforms usually have multiple applications with shared user accounts (e.g., search, news, and video recommendation in Google), which can meet users' diverse demands.
Users' various behaviors in multiple applications are related via shared accounts after user approvals, which can provide additional information from different aspects for the target recommendation task.
In real-world scenarios, recommendation systems benefit from not only effective algorithms, but also informative data. A good online recommendation should make full use of both \emph{user internal behaviors} (i.e., user behaviors in the target recommendation task) and \emph{user external behaviors} (i.e., user behaviors in other applications).
(2) \textbf{\emph{User preferences are variable}}. In large-scale recommendation systems, millions of new items are daily generated and added to the candidate pool. Online systems (especially news and video recommendations) should capture users' short-term preferences timely and accurately, since users' concentrations are easily attracted by new trends and hot topics.
In contrast, users' long-term preferences could provide users' aggregated stable interests as effective supplements to short-term interests.
Hence, a good real-world recommendation should well capture both \emph{user variable short-term preferences} and \emph{user stable long-term preferences}.

\begin{figure}[!hbtp]
\centering
\includegraphics[width=0.99\columnwidth]{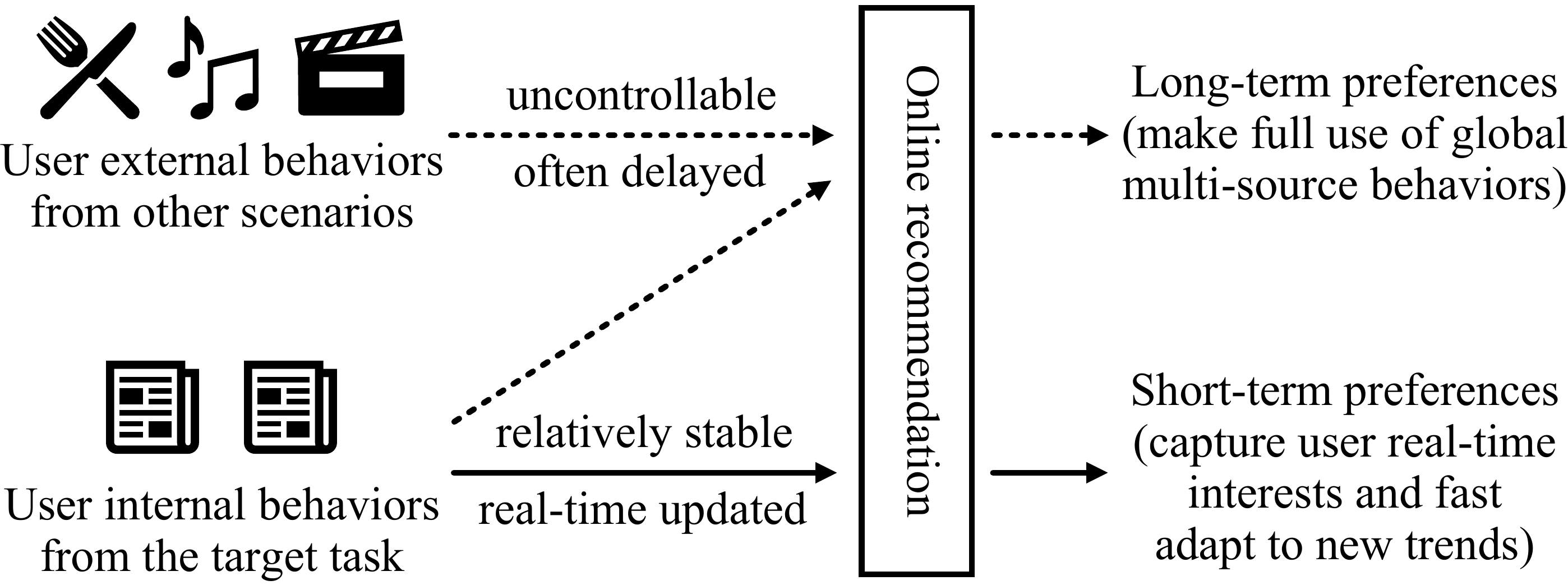}
\caption{A practical online recommendation with asynchronous multi-source internal and external behaviors, aiming to model users' short-term and long-term preferences.}
\label{fig:example}
\end{figure}

In this work, we attempt to design an effective and efficient online recommendation framework, which jointly considers both \textbf{users' internal/external behaviors} and \textbf{users' short-term/long-term preferences}. This online recommendation mainly faces the following three challenges:
(1) \emph{How to jointly consider internal and external behaviors?} As in Fig. \ref{fig:example}, users' multi-source behaviors in various applications usually have different features. It is difficult to combine this heterogeneous information. Moreover, in real-world scenarios, users' internal and external behaviors are often asynchronous (i.e., real-time internal feedbacks V.S. delayed and uncontrollable external behaviors), which makes it inconvenient to conduct a stable joint learning.
(2) \emph{How to effectively model both short-term and long-term preferences?} Both users' short-term and long-term interests are essential in real-world recommendations, while there are biases between them. Models should capture both short-term and long-term preferences, and should be able to determine which interest is more essential in different scenarios.
(3) \emph{How to conduct a timely online update to capture short-term preferences?} It is crucial to conduct an online update to capture user variable interests over time. However, it is extremely time-consuming (or even impractical) to conduct a complete model retraining or a complicated fine-tuning with million-level new behaviors. It is challenging to balance effectiveness and efficiency in online recommendation.
These three challenges are important in real-world scenarios, while there are barely any works that jointly address them systematically.

To address these issues, we propose a novel \textbf{Long Short-Term Temporal Meta-learning (LSTTM)} framework for practical online recommendations. Specifically, we build two heterogeneous graphs, namely the global long-term graph and the internal short-term graph, to capture different user preferences from multi-source behaviors.
The \emph{global long-term graph} is a huge graph containing all users' internal and external behaviors. It aims to build a global view of all multi-source interactions between users and items to better capture users' long-term preferences. In contrast, the \emph{internal short-term graph} focuses on the short-term behaviors of the target recommendation task, which is specially optimized for real-time interest evolutions. LSTTM adopts graph attention networks (GATs) with different neighbor sampling strategies to learn user long-term and short-term representations from these heterogeneous graphs, and then combines them via a gating fusion.
To better capture long-/short-term preferences and balance effectiveness and efficiency in online serving, we further design an asynchronous optimization method. We propose a \textbf{temporal MAML} training strategy, which \emph{regards recommendations at different time periods as different tasks}, based on a classical meta-learning method named MAML \cite{finn2017model}. This new temporal meta-learning enables fast adaptations to real-time variable user preferences.
The advantages of LSTTM mainly locate in three aspects:
(1) LSTTM makes full use of all internal and external behaviors via two huge graphs.
(2) We conduct different GAT aggregations and training strategies for two graphs to learn user short-term and long-term preferences separately, and combine them via gating. The differential designs enable more refined preference learning.
(3) The asynchronous optimization with temporal MAML facilitates fast adaptations to new trends under the practical (data and computation) limitations of real-world systems.

In experiments, we conduct an offline temporal CTR prediction with competitive baselines on a real-world recommendation system, and also conduct an online A/B test. The significant offline and online improvements show the effectiveness of LSTTM. Moreover, we also conduct an ablation study to better understand the effectiveness of different components. The contributions of this work are concluded in four points as follows:
\begin{itemize}
  \item We first systematically address the practical challenges of jointly considering users' internal/external behaviors and short-/long-term preferences in recommendation via our new proposed LSTTM framework. LSTTM is effective and easy to deploy in practical systems.
  \item We build two graphs focusing on different aspects to make full use of all internal/external behaviors. Moreover, we set customized GAT aggregators and training strategies to better learn user short-/long-term preferences.
  \item We design a novel temporal meta-learning method based on MAML, which enables fast adaptations to users' real-time preferences. To the best of our knowledge, we are the first to adopt temporal MAML in online recommendation.
  \item We achieve significant improvements on offline and online evaluations. LSTTM has been deployed on a real-world system for millions of users. The idea of temporal MAML can also be easily transferred to other models and tasks.
\end{itemize}

\section{Related Works}

\textbf{Recommender System.}
In real-world recommendation, Factorization machine (FM) \cite{rendle2010factorization}, NFM \cite{he2017neural}, DeepFM \cite{guo2017deepfm}, AutoInt \cite{song2019autoint}, DFN \cite{xie2020deep} are widely used to model feature interactions.
User behaviors are one of the most essential features to learn user preferences. Lots of models \cite{zhou2018deep,sun2019bert4rec,xie2020internal,xiao2021uprec,zeng2021knowledge} regard user behaviors as sequences to model user preferences via attention and transformer.
Besides sequence-based models, graph-based models such as SR-GNN \cite{wu2019session} and GCE-GNN \cite{wang2020global} use graph neural networks (GNNs) on user behavior graphs built from sessions.
Inspired by these works, we also adopt GAT \cite{velivckovic2018graph} to model user internal/external behaviors, and use DeepFM to model long-/short-term feature interactions.

Both long- and short- term preferences are essential in recommendation.
\citeauthor{xiang2010temporal} \shortcite{xiang2010temporal} proposes an injected preference fusion on a session-based temporal graph to model users' long-term and short-term preferences simultaneously.
STAMP \cite{liu2018stamp} highlights users' current interests from the short-term memory of the last clicks.
DIEN \cite{zhou2019deep} explicitly models user's interest evolutions.
Some works jointly consider short-term and long-term representations \cite{an2019neural,yu2019adaptive}.
\citeauthor{hu2020graph} \shortcite{hu2020graph} conducts LSTM for short-term interests and GNNs for long-term preferences. Intent preference decoupling \cite{liu2020intent} and online exploration \cite{xie2021hierarchical} are used for online recommendation.
However, these models cannot capture real-time global hot topics well, and do not make full use of multiple behaviors.
LSTTM builds two graphs focusing on long-term and short-term interests, and uses a temporal MAML to highlight short-term interests.

\textbf{Meta-learning in Recommendation.}
Meta-learning aims to transfer the meta knowledge so as to rapidly adapt to new tasks with a few examples, which is regarded as ``learning to learn'' \cite{vilalta2002perspective}.
MAML \cite{finn2017model} is a classical model-agnostic meta-learning method widely used in various fields, which provides a good weight initialization.
Some works have also explored meta-learning on graphs \cite{zugner2019adversarial,luo2020learning}.

In recommendation, meta-learning has also been verified on various cold-start scenarios, including cold-start user \cite{lee2019melu,zhu2021learning}, item \cite{vartak2017meta,zhu2021learn}, cross-domain recommendation \cite{du2019sequential,zhu2021transfer} and model selection \cite{luo2020metaselector}.
MeLU \cite{lee2019melu} brings in MAML to model cold-start users. \citeauthor{bharadhwaj2019meta} \shortcite{bharadhwaj2019meta} improves MAML with the dynamic meta-step for cold-start users.
\citeauthor{lu2020meta} \shortcite{lu2020meta} combines MAML with heterogeneous information networks, considering both semantic-wise and task-wise adaptations.
Besides MAML-based methods, \citeauthor{pan2019warm} \shortcite{pan2019warm} proposes meta-embeddings for warm-up scenarios.
MAMO \cite{dong2020mamo} designs two task-specific and feature-specific memories.
SML \cite{zhang2020retrain} focuses on model retraining, which learns a transfer function from old to new parameters.
Different from these models, LSTTM designs a temporal MAML to accelerate model adaptation to users' short-term preferences. To the best of our knowledge, we are the first to conduct temporal MAML in recommendation.

\section{Methodology}

In this work, we attempt to jointly consider both user internal and external behaviors to capture users' short-term/long-term preferences in online recommendation. We first give brief definitions of the notions used in this work as follows:

\noindent
\textbf{Definition 1:}
\textbf{User internal behaviors.} In LSTTM, the user behaviors of the target recommendation task are viewed as the user internal behaviors. These behaviors are the main sources of user preferences inside the target recommendation.

\noindent
\textbf{Definition 2:}
\textbf{User external behaviors.} All user behaviors from other applications are considered as the user external behaviors. These user external behaviors are related to their internal behaviors via the shared user accounts under user approvals. These behaviors are informative complements to the internal behaviors.

\noindent
\textbf{Definition 3:}
\textbf{Temporal meta-learning.} Meta-learning aims to fast adapt to new tasks \cite{finn2017model}. We define the temporal meta-learning, which \emph{regards recommendations in different time periods as different tasks}, since user preferences can frequently change over time. The temporal meta-learning concentrates on fast adaptations between user behaviors at different times for short-term preferences.

\subsection{Overall Framework}
\label{sec.overall_architecture}

\begin{figure}[!hbtp]
\centering
\includegraphics[width=0.92\columnwidth]{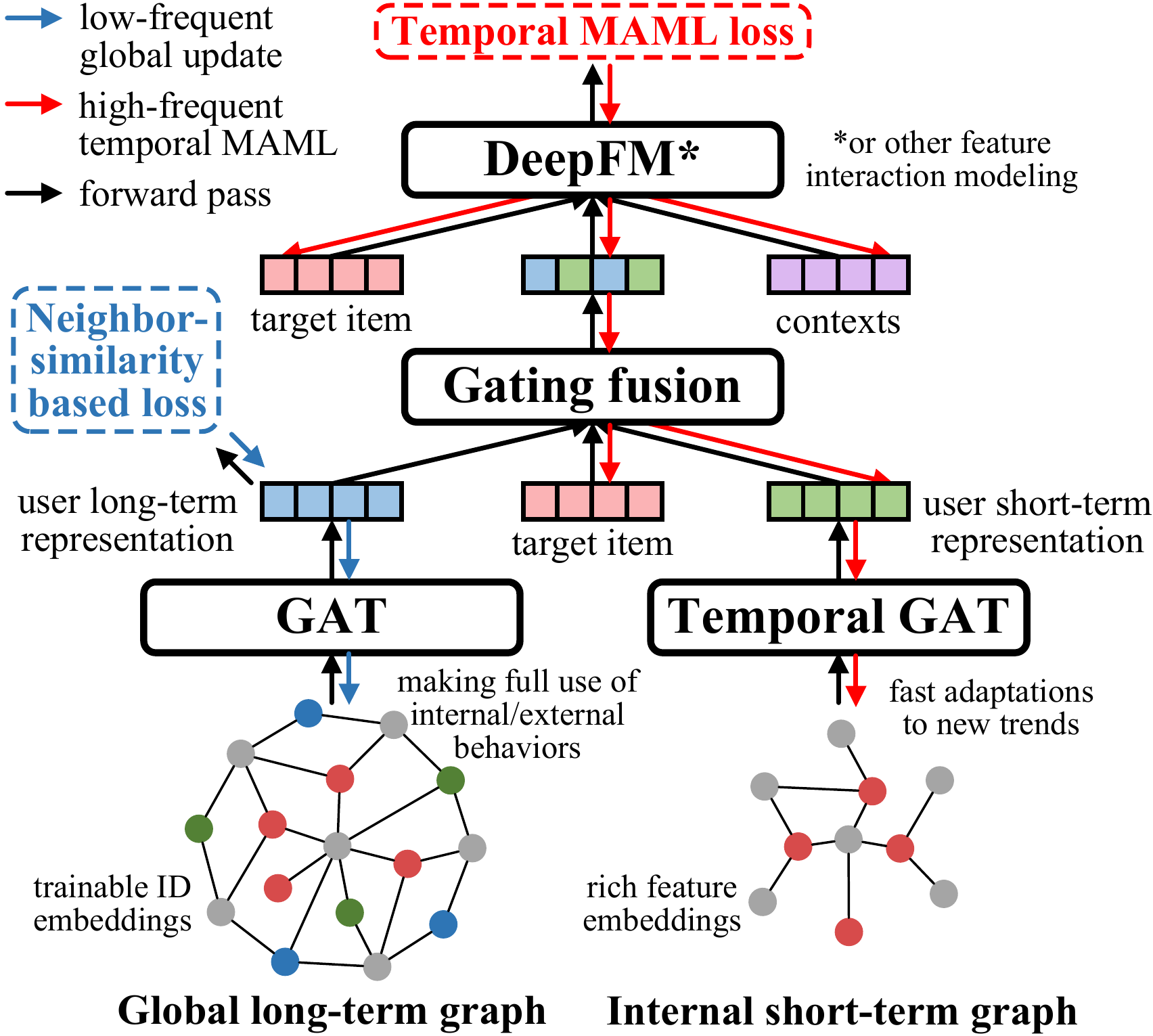}
\caption{Overall architecture of LSTTM.}
\label{fig:architecture}
\end{figure}

Fig. \ref{fig:architecture} displays the overall architecture of LSTTM.
The internal short-term graph regards all users and items in the target recommendation task as nodes, with all user internal behaviors utilized as edges. A heterogeneous GAT with temporal neighbor sampling is used for node aggregation to highlight user short-term preferences.
In contrast, the global long-term graph is a much larger heterogeneous graph having all user internal and external behaviors, which focuses on the global view of user long-term preferences in multi-source behaviors.
The long short-term graph fusion module then conducts a gating strategy to combine two short-/long-term representations, followed by a feature interaction module to generate the final recommendation.
We propose an asynchronous optimization to learn long-term and short-term preferences differently. For the short-term graph and gating fusion modules, we propose a temporal MAML to highlight short-term interest modeling. While for the long-term graph, we rely on a multi-hop neighbor-similarity based loss for efficient long-term preference learning.
In this case, LSTTM can make full use of both internal and external information to leverage user short-term and long-term preferences, finding an industrial balance between effectiveness and efficiency in practice.

\subsection{Internal Short-term Graph}
\label{sec.internal_graph}

The internal short-term graph module models user short-term internal behaviors. Specifically, it has two types of nodes indicating all users $u \in U$ and all items $d \in D_I$ in the target recommendation task, which are linked by user-item click behaviors as edges. $U$ and $D_I$ are the overall user set and internal item set respectively.
We use $\bm{u}_i^0$ and $\bm{d}_i^0$ to represent the $i$-th input feature embeddings of users and items, which are trainable embeddings built from different types of user attributes and item features (e.g., tag, topic).

Inspired by \citeauthor{velivckovic2018graph} \shortcite{velivckovic2018graph}, we build an enhanced GAT layer for short-term oriented node aggregation. For a user $u_i$ and his/her click sequence $seq_i=\{d_{i,1}, \cdots, d_{i,m}\}$, different from conventional random-based neighbor sampling, we conduct a temporal neighbor sampling, which only selects the top-K most recent clicked items. The temporal neighbor sampling generates the neighbor set $N_{u_i}$ as:
\begin{equation}
\begin{split}
N_{u_i} = \mathrm{Temporal}(seq_i) = \{d_{i,m-K+1}, \cdots, d_{i,m}\}.
\label{eq.short_neighbor}
\end{split}
\end{equation}
Similarly, we also generate the sampled neighbor set of items as $N_{d_i} = \{u_{i,m'-K+1}, \cdots, u_{i,m'}\}$. With the temporal neighbor set $N_{u_i}$, we build the user representation $\bm{u}_i^k$ at the $k$-th layer via item embeddings in the $k-1$ layer as follows:
\begin{equation}
\begin{split}
\bm{u}_i^k = \sigma (\sum_{d_{i,j} \in N_{u_i}} \alpha^k_{ij} \bm{W}^{k}_d \bm{d}^{k-1}_{i,j} ).
\label{eq.aggregation}
\end{split}
\end{equation}
$\bm{W}^{k}_d$ is the weighting matrix. $\alpha^k_{ij}$ represents the attention between $u_i$ and $d_{i,j}$ in this layer, which is formalized as:
\begin{equation}
\begin{split}
\alpha^{k}_{ij} = \frac {\exp ( f( \bm{w}^{\top} [\bm{W}_u \bm{u}^{k-1}_i || \bm{W}_d \bm{d}^{k-1}_{i,j} ]))}
{\sum_{d_{i,l} \in N_{u_i}} \exp ( f( \bm{w}^{\top} [\bm{W}_u \bm{u}^{k-1}_i || \bm{W}_d \bm{d}^{k-1}_{i,l} ]))},
\label{eq.short_attention}
\end{split}
\end{equation}
where $f(\cdot)$ indicates a LeakyReLU activation and $||$ indicates the concatenation. $\bm{w}^{\top}$, $\bm{W}_u$ and $\bm{W}_d$ are the weighting vector and matrices. Note that the temporal neighbor set $N_{u_i}$ changes over time, since the internal short-term graph is a dynamic graph that is updated via users' new behaviors. Finally, we conduct a two-layer temporal GAT to generate the user short-term representation $\bm{u}^s_i=\bm{u}^2_i$, which is fed into the next gating fusion module. The aggregation of items is similar to that of users. We use GAT since it is effective, efficient, and easy to deploy on billion-level huge graphs. It is also convenient to conduct other enhanced GNN models in this module.

In internal short-term graph, the temporal neighbor sampling highlights the individual-level short-term preferences. We also propose a temporal meta-learning method to update this module, attempting to capture the short-term preferences at the global level, which will be introduced in Sec. \ref{sec.optimization_objective}.

\subsection{Global Long-term Graph}
\label{sec.global_graph}

The global long-term graph module aims to take advantage of all user diverse preferences in multiple applications. It considers users $u \in U$ and all internal and external items $d \in D_I \bigcup D_E$ as nodes, where $D_I$ and $D_E$ represent the overall internal and external item sets. All heterogeneous user-item interactions in different applications as regarded as edges. The details of the external behaviors are in Sec. \ref{sec.online_deployment}. Since heterogeneous items usually have different feature fields that are hard to align, we represent all users and items via trainable ID embeddings $\bm{\bar{u}}_i^0$ and $\bm{\bar{d}}_i^0$ in the same space.

We also conduct a two-layer GAT for neighbor aggregation similar as Eq. (\ref{eq.aggregation}) to Eq. (\ref{eq.short_attention}), where the neighbor set $\bar{N}_{u_i}$ are randomly sampled or selected via certain importances.
The user long-term representation $\bm{\bar{u}}^l_i = \bm{\bar{u}}_i^2$ is also utilized in the gating fusion module. Since the overall behaviors are too enormous to be fully retrained in online, and external behaviors are usually delayed and uncontrollable, we conduct an enhanced neighbor-similarity based loss to train this module asynchronously introduced in Sec. \ref{sec.optimization_objective}.

Comparing the internal short-term graph modeling with the global long-term graph modeling, we can find three main differences:
(1) they adopt different data sources, and thus have different input feature forms (i.e., detailed user and item features V.S. trainable ID embeddings).
(2) They conduct different neighbor sampling strategies (i.e., temporal-based V.S. random or importance-based) due to their different long-/short- term concentrations.
(3) They are updated under different strategies (i.e., temporal meta-learning V.S. neighbor-similarity based loss), considering the effectiveness of short-term preference modeling and the efficiency of global user behavior modeling in practice.
We use GNN to capture heterogeneous node interactions, and conduct GAT in node aggregation for efficiency. It is also not difficult to conduct other complicated heterogeneous information networks in LSTTM.

\subsection{Long- Short-term Preference Fusion}
\label{sec.graph_fusion}

This module attempts to combine both user short-term and long-term representations $\bm{u}_i^s$ and $\bm{\bar{u}}_i^l$ to generate the ranking score. We conduct a gating-based fusion to generate the final user representation $\bm{u}_i$ via $\bm{u}_i^s$ and $\bm{\bar{u}}_i^l$ as follows:
\begin{equation}
\begin{split}
\bm{u}_i = g(\bm{x}_i^s) \bm{u}_i^s + g(\bm{x}_i^l) \bm{\bar{u}}_i^l.
\end{split}
\end{equation}
$g(\cdot)$ indicates the gating function, which is measured via the corresponding user embeddings and the target item $\bm{d}_j^s$ as:
\begin{equation}
\begin{split}
[g(\bm{x}_i^s),  g(\bm{x}_i^l)] = \mathrm{Softmax} ([\bm{w}^s_g [\bm{u}_i^s || \bm{d}_j^s], \bm{w}^l_g [\bm{\bar{u}}_i^l || \bm{d}_j^s]]).
\end{split}
\end{equation}
$\bm{w}^s_g$ and $\bm{w}^l_g$ are weighting vectors. $\bm{d}_j^s$ is a trainable item ID embedding that is randomly initialized. With this gating-based fusion, users can get personalized weights on long-/short- term preferences for different items, which helps to improve the performances.

After gating fusion, the final user representation $\bm{u}_i$ is aggregated with the recommendation contexts $\bm{c}$ and target item embedding $\bm{d}_j^s$, and then fed into the downstream neural ranking models. We conduct a widely-used DeepFM \cite{guo2017deepfm} to model the feature field interactions between user, item and contexts as follows:
\begin{equation}
\begin{split}
p(i,j) = \textrm{DeepFM} (\bm{u}_i, \bm{d}_j^s, \bm{c}).
\end{split}
\end{equation}
$p(i,j)$ is the click probability for $u_i$ and $d_j$. It is also easy to adopt other feature interaction models for feature interactions.

\subsection{Optimization with Temporal MAML}
\label{sec.optimization_objective}

The asynchronous optimization with temporal MAML is the key contribution of LSTTM.
In practice, timely model updating is significant in online recommendation, while there are two challenges in real-world systems. (1) It is extremely difficult to conduct a full model retraining or a complicated fine-tuning in real-time for GNN models with large-scale graphs, especially with the nearly billion-level interactions in the huge global graph. (2) Moreover, multi-source behaviors are usually obtained asynchronously (e.g., external behaviors are often delayed) due to some practical system limitations. Hence, we decouple the training of internal short-term graph and global long-term graph into two asynchronous optimization objectives, including a temporal MAML based cross-entropy loss and a multi-hop neighbor-similarity based loss. It enables LSTTM to be smoothly and timely updated.

\subsubsection{Temporal Meta-learning}
\label{sec.temporal_MAML}

To enhance LSTTM with the capability of fast adaptation to user short-term interests, we propose a novel \textbf{temporal MAML} training strategy based on \cite{finn2017model}. Different from conventional meta-learning based recommendations that usually consider each user or domain as a task, our temporal MAML \textbf{regards recommendation in each time period as a task}.

Specifically, we first divide all training instances into different sets according to their time periods (e.g., we view each hour as a time period for the practical demands). In temporal MAML training, we regard instances in two adjacent hours as a task. The \emph{support set} contains instances of the former hour, while the \emph{query set} contains instances of the latter hour. Note that an instance can belong to both a support set and a query set in two tasks.
We could further divide these temporal tasks into more fine-granularity tasks, where all instances in a task derive from the same user group (which is built via similar basic profiles or user interests) at the same time. Through these fine-granularity tasks, the instances in the support set and the query set will be more relevant. In this case, the temporal MAML will focus on the new trends in certain user communities instead of whole user groups. We can choose different temporal MAML settings according to the practical needs of systems.

In training, we sample different temporal tasks containing training instances in different time periods to form a batch. To make sure our temporal MAML can learn a better initialization for fast adaptation on all time periods, we diversify the sampled temporal tasks to make them dissimilar to each other (e.g., selecting tasks in different hours and days). The inner update (line $6$) with support sets and the outer update (line $8$) with query sets are similar to the original MAML. We conduct one gradient update in both inner and outer updates for efficiency.
In online, we also conduct one gradient update for all new user feedbacks similar to MAML, which enables high-frequent (or even real-time) online model learning.
Algorithm \ref{algorithm1} gives the pseudo-code of temporal MAML with temporal tasks.

\begin{algorithm}[!htbp]
\small
  \caption{\textbf{Temporal MAML:}}
  \label{algorithm1}
  \begin{algorithmic}[1]
    \Require
      The distribution over temporal tasks $p(T)$
    \Ensure
      The parameter set $\theta=(\theta_s,\theta_f)$ of the internal short-term graph module $\theta_s$ and the long- short-term preference fusion $\theta_f$
    \State Randomly initialize $\theta$
    \While{not converge}
        \State Sample batch of diversified temporal tasks $T_i \sim p(T)$
        \ForAll {$T_i$}
            \State Evaluate $\nabla_\theta L_{T_i} (f_\theta)$ with respect to $K$ examples in each
            \Statex \quad \quad \quad support set, using the graphs in $T_i$'s time period
            \State Compute adapted parameters with gradient descent via
            \Statex \quad \quad \quad the inner step size $\alpha$: $\theta'_i = \theta - \alpha \nabla_\theta L_{T_i} (f_\theta)$
        \EndFor
        \State Update the parameter $\theta$ via the outer step size $\beta$: $\theta \leftarrow \theta -$
         \Statex \quad \ \ $\beta \nabla_\theta \sum_{T_i \sim p(T)}  L_{T_i} (f_{\theta'_i})$ with all query sets
    \EndWhile
  \end{algorithmic}
\end{algorithm}

Under the temporal MAML training framework, we conduct a classical cross entropy loss $L_T$ with the click probability $p(i,j)$ of user $u_i$ and item $d_j$ on the positive set (clicked user-item instances) $S_p$ and negative set (unclicked user-item instances) $S_n$ as follows:
\begin{equation}
\begin{split}
L_T=-\frac{1}{N}(\sum_{S_p} \log p(i,j) + \sum_{S_n} \log (1-p(i,j)) ).
\label{eq:cross_entropy}
\end{split}
\end{equation}
Note that the $L_T$ is only used for updating the internal short-term graph and the long- short-term preference fusion modules via the temporal MAML as in Fig. \ref{fig:architecture}.
A gradient block is conducted to the global long-term graph module, since it is responsible for modeling users' stable long-term preferences from multi-source behaviors, and thus should be fully trained on all behaviors.

\noindent
\textbf{Motivations and advantages of temporal MAML.}
The motivations and advantages of the temporal MAML are concluded as follows:
(1) we attempt to capture new trends and hot topics timely in practical recency-sensitive recommendation systems. The temporal MAML highlights the model's capability in capturing global user interest evolutions via temporal tasks, which enables fast adaptations to users' variable short-term interests on the global new trends.
(2) MAML can fast adapt to new tasks \cite{finn2017model}, while classical MAML-based models mainly regard individual users as tasks, and thus cannot model the global temporal factors well. Hence, we propose the temporal MAML focusing on the temporal tasks.
(3) The internal short-term graph modeling can also provide short-term interests via the temporal neighbor sampling (see Eq. (\ref{eq.short_neighbor})). However, it merely concentrates on the individual user-related short-term behaviors, ignoring the global short-term behaviors generated by other users (which is essential especially when the user does not have recent behaviors). The temporal MAML and the short-term graph modeling are strong supplements to each other in capturing user real-time preferences.
(4) The temporal MAML is also naturally suitable for the asynchronous online learning with large-scale instances, since it only needs a one-step update.

\subsubsection{Multi-hop Neighbor-similarity Based Loss}

Differing from the internal short-term graph, the global long-term graph (1) aims to model user long-term behaviors, (2) contains far more internal and external behaviors, and (3) might have uncontrollable and delayed behavior acquisitions. To make a compromise between efficiency, effectiveness, and robustness, we conduct a multi-hop neighbor-similarity based loss instead of the online temporal MAML.

We assume that both users' and items' long-term representations $\bm{\bar{u}}_i^l$ and $\bm{\bar{d}}_j^l$ learned in Sec. \ref{sec.global_graph} should be similar to their k-hop neighbors on the global long-term graph enhanced from \cite{liu2020graph} and \cite{xie2021improving}. The multi-hop neighbor-similarity based loss on the global user-item graph can be viewed as an extended matrix factorization (MF) model, which considers multi-hop user-item paths on the global graph as multi-source user/item correlations.
Precisely, we consider the 10-hop neighbors via DeepWalk based path sampling \cite{perozzi2014deepwalk} to bring in more interactions via users' multi-source behaviors. We formalize our k-hop neighbor-similarity based loss $L_N$ as follows:
\begin{equation}
\begin{split}
L_N = - \sum_{p \in P} \sum_{q_i, q_j \in p} (\log (\sigma(\bm{\bar{q}}_i^{l\top} \bm{\bar{q}}_j^l))).
\label{eq:neighbor_similarity_loss}
\end{split}
\end{equation}
$p$ is a k-length random path in the path set $P$ generated by DeepWalk. $q_i, q_j \in p$ are different nodes in the path $p$. We have $\bm{\bar{q}}_i^l=\bm{\bar{u}}_i^l$ for user nodes and $\bm{\bar{q}}_j^l=\bm{\bar{d}}_j^l$ for item nodes. $\sigma$ is the sigmoid function.
The multi-hop neighbor-similarity based loss focuses more on the global view of user and item representations learned from all long-term internal/external behaviors. Generally, the global long-term graph trains far less frequently than the short-term graph considering its motivation and training efficiency.

The advantages of using the multi-hop neighbor-similarity based loss for the global long-term graph are as follows:
(1) the $L_N$ loss is simple, efficient, and effective, which can directly optimize the cross-source interactions via the multi-hop connections.
(2) Based on the neighbor-similarity based loss, it is more convenient to introduce other heterogeneous nodes (e.g., content or tag in \cite{xie2021improving}) and their interactions in this work. Other node representation learning methods are also easy to be adopted in our framework.

\subsubsection{Overall loss}

The overall loss $L$ is the weighted aggregation of these two losses $L_T$ and $L_N$ as follows:
\begin{equation}
\begin{split}
L = \lambda_T L_T + \lambda_N L_N.
\label{eq:overall_loss}
\end{split}
\end{equation}
We empirically set $\lambda_T=\lambda_N=1$. In LSTTM, the neighbor-similarity based loss $L_T$ works as an auxiliary task for the temporal MAML loss $L_T$, since $L_T$ is directly related to the ranking objectives.

The advantages of our asynchronous optimization are listed as follows:
(1) it decouples the short-term and long-term preference modeling, making both modules more flexible, specialized, and robust to capture different user preferences.
(2) It proposes an industrial solution to jointly consider large-scale external and internal behaviors, improving the robustness against the high disturbances and uncontrollability in practical systems (e.g., the delay and noises of external behaviors have little influence on the internal short-term preference modeling).
(3) The asynchronous optimization is flexible and easy to deploy. It is also convenient to use other representation learning models in this asynchronous framework.

\section{Online Deployment}
\label{sec.online_deployment}

\noindent
\textbf{Online System.}
We have deployed LSTTM on a real-world recommendation system of WeChat. This online recommendation system is a feed stream that has nearly million-level users and daily views. It contains heterogeneous domains, including news and videos. User behaviors of the target recommendation task are viewed as the internal behaviors. After user approvals, other behaviors (e.g., clicks in other recommendation domains) linked by the same user account in the platform are regarded as the external behaviors, which also provide additional information to reflect user preferences.

\noindent
\textbf{Online Serving.}
We conduct an asynchronous optimization and online updating for different modules.
In offline training, we conduct a daily complete training to update all modules via the asynchronous optimization of temporal MAML and neighbor-similarity based losses on all behaviors. It constructs an industrial balance between effectiveness and efficiency, since the access to user external behaviors is usually delayed, and the full training on billion-level global graphs cannot be conducted in real-time.
In online serving, the global graph module is fixed within a day for modeling long-term preferences, while the internal graph and fusion modules are frequently updated (according to the online computing capability) to capture user short-term preferences. When recommending at the $t$-th time period, we consider \emph{all previous $t-1$ time periods} in this day as the support set, simulating the offline temporal MAML training. Hence, we \emph{just need to conduct the general one-step gradient updates on new user behaviors}, regarding them as the support set of the current recommendation. It enables a fast online learning since the online time complexity of temporal MAML is equivalent to the classical one-step fine-tuning.

\noindent
\textbf{Online Efficiency.}
We train our model once over all training instances in online updating considering the efficiency. The online computation does not involve the global long-term graph modeling. The online time complexity of LSTTM is $O(k(T_i+T_f))$, where $k$ is the number of candidates (e.g., top $200$ items retrieved by the previous matching module). $T_i$ and $T_f$ represent the computation costs of the internal short-term graph (2-layer GAT with the dynamic temporal neighbors) and the fusion.
For the online memory cost, the model should store the temporal MAML model and the fixed user long-term representations.
Specifically, we implement LSTTM on a self-developed distributed deep learning framework. We have $30$ parameter servers and $20$ workers for training. Each server has $10$G memory with $3$ CPUs, and each worker has $10$G memory with $5$ CPUs. We spend nearly $4$ hours for daily complete retraining.

\section{Experiments}

In this section, we conduct experiments to answer the following research questions:
(\textbf{RQ1}): How does LSTTM perform in offline temporal CTR prediction that simulates practical scenarios (Sec. \ref{sec.offline_evaluation})?
(\textbf{RQ2}): How does LSTTM perform in online A/B tests (Sec. \ref{sec.online_evaluation})?
(\textbf{RQ3}): What are the effects of different components (Sec. \ref{sec.ablation})?

\subsection{Dataset}
\label{sec.dataset}

Since there is no large-scale real-world dataset that contains both hourly-updated hot spots and user external behaviors, we build a new dataset NewsRec-21B extracted from a widely-used news recommendation system in WeChat.
Precisely, we randomly select $58$ million users and get nearly $1$ billion user internal behaviors with timestamps in the target news domain. We also use these users' $20$ billion external click behaviors from other recommendation domains in the same platform after user approval to build the global long-term graph. These internal and external behaviors are in the same platform, which are linked via the shared user accounts. All data are preprocessed via data masking to protect user privacy. The instances in the former eight days are regarded as the train set, and the last day's internal behaviors are considered as the test set. We follow Sec. \ref{sec.internal_graph} and Sec. \ref{sec.global_graph} to build two huge graphs with the train set.
Table \ref{tab:dataset} shows the detailed statistics of NewsRec-21B.

\begin{table}[!hbtp]
\centering
\small
\begin{tabular}{cccc}
\toprule
\#user & \#item & \#internal & \#external  \\
\midrule
 58,284,406 & 626,736 & 1,022,589,888 & 20,087,883,624  \\
\bottomrule
\end{tabular}
\caption{Statistics of the NewsRec-21B dataset.}
\label{tab:dataset}
\end{table}

\subsection{Competitors}
\label{sec.baseline}

We implement several competitive baselines for evaluation. First, we conduct four widely-used ranking models as follows:
\begin{itemize}
  \item \textbf{FM \cite{rendle2010factorization}.} FM is a simple and effective model that captures second-order feature interactions via latent vectors.
  \item \textbf{NFM \cite{he2017neural}.} NFM combines the neural FM layer with the DNN layer sequentially to model high-order feature interactions.
  \item \textbf{DeepFM \cite{guo2017deepfm}.} DeepFM follows the Wide\&Deep framework and improves the Wide part with a neural FM layer. It is also used in the long-/short- term gating fusion of LSTTM.
  \item \textbf{AutoInt \cite{song2019autoint}.} AutoInt is a strong feature interaction modeling method, which adopts self-attention layers.
\end{itemize}
These baselines use the same features of the users, internal behaviors and contexts that are also used in LSTTM, and are optimized via the same training set with the cross-entropy loss.

For fair comparisons, we also implement two enhanced DeepFM models armed with external behaviors and sequence modeling.
\begin{itemize}
  \item \textbf{DeepFM (+external).} We add the features of user external behaviors to DeepFM, noted as DeepFM (+external). It has the same input features as the LSTTM model.
  \item \textbf{DIN+DeepFM (+external).} Based on DeepFM (+external), we further bring in the ability of sequence-based modeling on user's internal and external behaviors to better model the short-term and long-term preferences. Inspired by \cite{xie2020deep}, we conduct two DIN encoders \cite{zhou2018deep} to model internal and external behaviors respectively. These behavior features are considered as the input feature fields of DeepFM (+external).
\end{itemize}

Finally, since we conduct the temporal MAML for online updating, we also implement two SOTA meta-learning methods based on SML \cite{zhang2020retrain} in online news recommendation as follows:
\begin{itemize}
  \item \textbf{SML \cite{zhang2020retrain}.} SML is the SOTA meta-learning based recommendation model designed for model retraining verified in online new recommendation \cite{zhang2020retrain}. It is the most related baseline of our task. SML attempts to learn a transfer function from old to new parameters via a sequential training over time. Following the original SML's implementation, we also build SML based on an MF model.
  \item \textbf{SML+DeepFM.} We further improve the original SML by replacing the MF model with the best performing DeepFM model, noted as SML+DeepFM. This model also utilizes the same features as LSTTM for fair comparisons.
\end{itemize}
Note that we do not compare with other meta-learning recommendation methods such as MeLU \cite{lee2019melu}, since they focus on different tasks (e.g., cold-start users or domains) and are not suitable for our temporal setting. To further verify the effectiveness of different components and features in LSTTM, we implement four ablation versions of LSTTM, whose results are discussed in Sec. \ref{sec.ablation}.

\subsection{Experimental Settings}
\label{sec.experimental_settings}

We randomly select up to $30$ neighbors in global graph and $30$ most recent behaviors in internal graph for aggregation. The dimensions of the output embeddings in both graphs are $16$. We use $6$ fields for users (e.g., user profiles such as age and gender) and items (e.g., item features such as tag and provider), and the dimension of each trainable feature field embedding is $16$.
In temporal MAML, the task number of each batch, batch size, and learning rate are essential parameters. We have tested the task number among $\{4,8,16\}$, the support and query set size among $\{32,64,128,256\}$, and the learning rate among $\{0.001, 0.01, 0.02\}$. Finally, we let each batch size contain $8$ temporal tasks of different days and hours, where each support set and query set contain $128$ items. We use Adagrad \cite{duchi2011adaptive} and empirically set the same learning rate as $0.01$ for inner and outer updates. Note that all instances could belong to a certain query set used in Line 8, Algorithm \ref{algorithm1}, which directly updates model parameters.
We only conduct a one-step gradient in temporal MAML for online efficiency. We conduct a grid search for parameter selection. All models share the same experimental settings.

\subsection{Temporal CTR Prediction (RQ1)}
\label{sec.offline_evaluation}

We first simulate the real-world online recommendation and conduct the temporal CTR prediction task for offline evaluation.

\subsubsection{Evaluation Protocol}

We evaluate models on our real-world dataset NewsRec-21B. To simulate the online recommendation, we first train all models with the train set (the former few days), and divide the test set (the last day) into $24$ hours. Each hour is regarded as a temporal task for evaluation, with all instances of former hours in the test set used as the support set. Considering both accuracy and online efficiency, all models including LSTTM and baselines are fine-tuned via one gradient update for fair comparisons (SML is updated via its transfer method \cite{zhang2020retrain}).
We use AUC as our metric, which is widely utilized as the main metric in real-world systems \cite{guo2017deepfm,zhou2018deep,song2019autoint}. For a better display, we group $24$ hours into $3$ periods (0:00-8:00, 8:00-16:00, 16:00-0:00+1), and report the average AUC in each period in Table \ref{tab:CTR}. We conduct $3$ runs for each model.

\begin{table}[!hbtp]
\centering
\small
\begin{tabular}{l|c|c|c}
\toprule
AUC & period1 & period2 & period3  \\
\midrule
FM (\citeauthor{rendle2010factorization} \citeyear{rendle2010factorization}) & 0.8086 & 0.8021 & 0.8072 \\
NFM (\citeauthor{he2017neural} \citeyear{he2017neural}) & 0.8140 & 0.8035 & 0.8154 \\
DeepFM (\citeauthor{guo2017deepfm} \citeyear{guo2017deepfm}) & 0.8253 & 0.8306 & 0.8296 \\
AutoInt (\citeauthor{song2019autoint} \citeyear{song2019autoint}) & 0.8236 & 0.8282 & 0.8272 \\
\midrule
DeepFM (+external) & 0.8282 & 0.8335 & 0.8327 \\
DIN+DeepFM (+external) & \underline{0.8287} & \underline{0.8346} & \underline{0.8337} \\
\midrule
SML (\citeauthor{zhang2020retrain} \citeyear{zhang2020retrain}) & 0.7926 & 0.7991 & 0.8063 \\
SML+DeepFM  & 0.8250 & 0.8307 & 0.8299 \\
\midrule
LSTTM (w/o Meta) & 0.8373 & 0.8428 & 0.8421 \\
LSTTM (final) & \textbf{0.8395} & \textbf{0.8504} & \textbf{0.8502} \\
\bottomrule
\end{tabular}
\caption{Results of the temporal CTR prediction task. The improvements of LSTTM are significant (t-test with p<0.01). Note that period1 focuses on the performances of models in a short time after complete training, while period3 focuses on the performances in a long time after complete training.}
\label{tab:CTR}
\end{table}

\subsubsection{Experimental Results}

\begin{figure}[!hbtp]
\centering
\includegraphics[width=0.99\columnwidth]{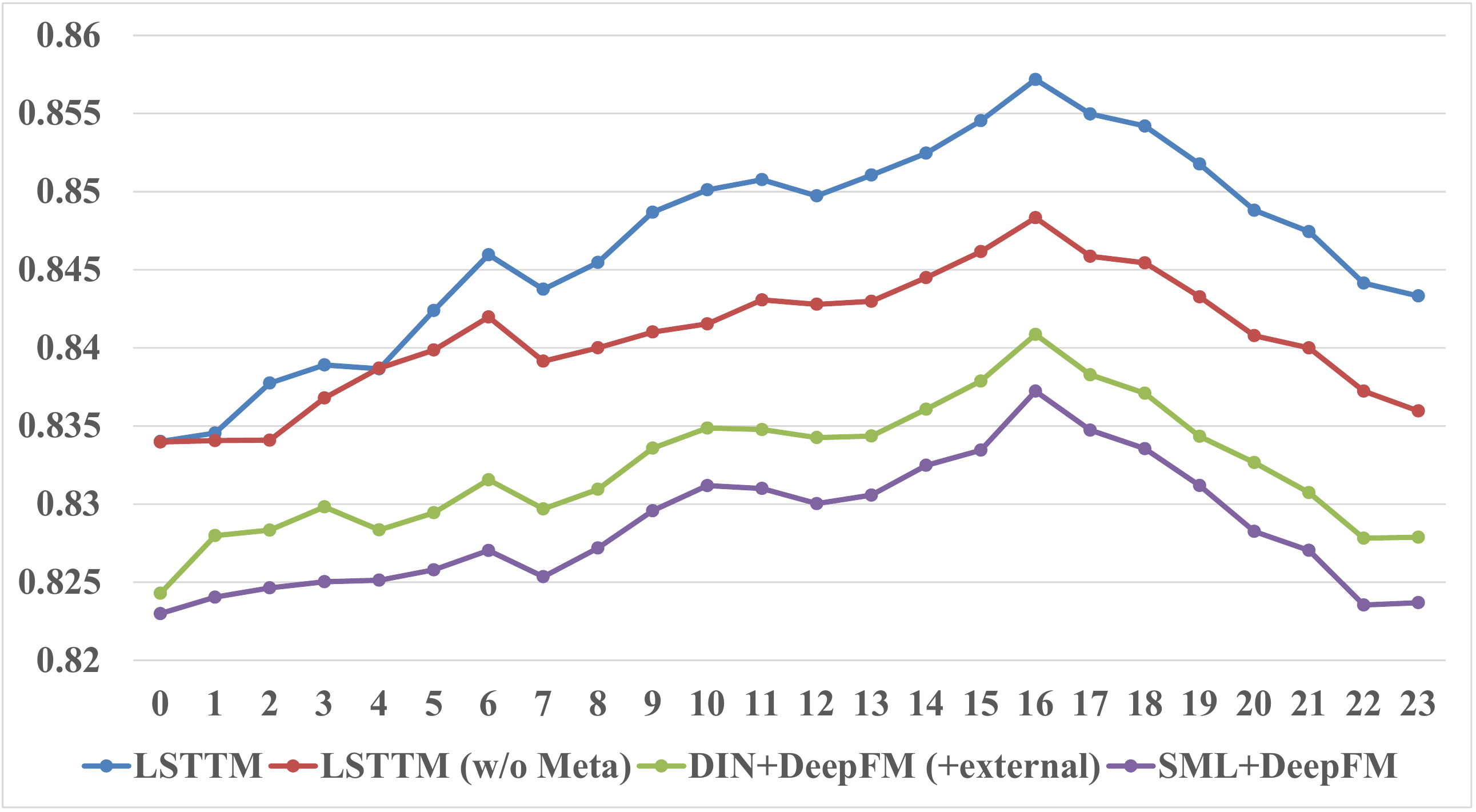}
\caption{AUC trends in different hours.}
\label{fig:hourly_results}
\end{figure}

From Table \ref{tab:CTR} we can observe that:

(1) LSTTM achieves significant improvements on all baselines in three periods, with the significance level $\alpha=0.01$. It consistently outperforms strong baselines in all $24$ hours (see Fig. \ref{fig:hourly_results}). The deviation is less than $\pm0.002$. Considering the large size of our test set, the $1.1\%-1.7\%$ AUC improvements over the best baseline are impressive and solid. It verifies the effectiveness and robustness of LSTTM in modeling both short-term and long-term preferences from users' internal and external behaviors.

(2) LSTTM (final) consistently outperforms LSTTM (w/o Meta) and SML on all tasks. It confirms the advantages of temporal MAML in Sec. \ref{sec.temporal_MAML}. Thanks to the MAML-based training, LSTTM is more sensitive to global new trends in communities. Hence, it can better capture users' short-term preferences via good model initialization, and thus can fast adapt to hot topics over time in online recommendation.
Nevertheless, LSTTM (w/o Meta) still performs better than baselines, which reflects the effectiveness of our global long-term and internal short-term graphs as well as the gating fusion. Sec. \ref{sec.ablation} gives more details of different ablation versions.

(3) We also find that models armed with external behaviors consistently outperform the same models without external behaviors (e.g., see LSTTM in Sec. \ref{sec.ablation} and DeepFM in Table \ref{tab:CTR}). It verifies the importance of external behaviors in real-world scenarios, which works as a strong supplement to the internal behaviors. The external behaviors will be more significant in few-shot scenarios.

(4) Comparing models in different periods, we know that LSTTM achieves larger improvements in period 2 and 3 compared to LSTTM (w/o Meta). It is because that (a) humans and hot spots are often more active in period 2 and 3, where temporal MAML is superior to baselines in capturing user real-time preferences. (b) In period3, all models have not been fully trained for at least 16 hours. LSTTM has a better online fine-tuning to catch up with new global interest evolutions. The cumulative effects of temporal MAML will gradually show up over time with growing hot topics. Fig. \ref{fig:hourly_results} shows the hour-level AUC trends of four representative models.

\subsection{Online A/B Tests (RQ2)}
\label{sec.online_evaluation}

\subsubsection{Evaluation Protocol}

To evaluate LSTTM in real-world systems, we conduct an online A/B test on WeChat Top Stories. Following Sec. \ref{sec.online_deployment}, we deploy our LSTTM in the ranking module of the news domain with other modules unchanged. The online base model is DeepFM with the same online update frequency of LSTTM. In online evaluation, we focus on four representative metrics, including Click-through-rate (CTR), average click number per capita (ACN), user has-click rate (HCR), and average dwell time per capita (DT) to measure recommendation accuracy and user satisfaction, which are formalized as follows:
\begin{equation}
\begin{split}
&\mathrm{CTR}=\frac{\mathrm{\#\ of\ all\ clicks}}{\mathrm{\#\ of\ all\ impressions}},\quad \
\mathrm{ACN}=\frac{\mathrm{\#\ of\ all\ clicks}}{\mathrm{\#\ of\ all\ users}},\\
&\mathrm{HCR}=\frac{\mathrm{\# \ of\ users\ having\ clicks}}{\mathrm{\#\ of\ all\ users}},\ \
\mathrm{DT}=\frac{\mathrm{all\ duration}}{\mathrm{\#\ of\ all\ users}}.
\end{split}
\end{equation}
We conduct the A/B test for $5$ days.

\subsubsection{Experimental Results}

Table \ref{tab:online} shows the improvement percentages over the base model. We can observe that:

(1) LSTTM achieves significant improvements on all metrics with the significance level $\alpha=0.01$. It reconfirms the effectiveness of LSTTM in online.
Through the asynchronous online updating with the temporal MAML, LSTTM can (a) fast adapt to new topics and hot spots, and (b) successfully combine both external and internal behaviors in online ranking without many computation costs.

(2) The improvement on CTR indicates that more appropriate items have been impressed to users (reflecting item-aspect accuracy), while the improvement on ACN represents that users are more willing to click items (reflecting user-aspect accuracy and activeness). HCR models the coverage of users that have clicked news, which implies the impacts of our recommendation function. LSTTM also outperforms the online baseline on dwell time of items, which reflects the real user satisfaction on the item contents. In conclusion, LSTTM achieves comprehensive improvements on all online metrics, which confirms the robustness of our model.

\begin{table}[!hbtp]
\centering
\small
\begin{tabular}{l|c|c|c|c}
\toprule
metrics & CTR & ACN & HCR & DT  \\
\midrule
LSTTM & +9.60\% & +9.93\% & +3.42\% & +2.52\% \\
\bottomrule
\end{tabular}
\caption{Online A/B tests on a widely-used system.}
\label{tab:online}
\end{table}

\subsection{Ablation Tests (RQ3)}
\label{sec.ablation}

We further conduct an ablation test to verify the effectiveness of different components in LSTTM. Table \ref{tab:ablation_test} shows the results of different ablation settings. We observe that all components significantly benefit the recommendation. Precisely, we find that:

(1) the temporal MAML can precisely capture user's variable short-term interests without additional online computation costs. The improvements are larger when more new trends are involved as time passes by, such as in period 2 and 3.

(2) The second ablation version only considers internal behaviors by removing the global long-term graph. It verifies the effectiveness of the user external behaviors as well as the global long-term graph modeling in Sec. \ref{sec.global_graph}. The advantages of external behaviors will be more significant if we deploy LSTTM on cold-start scenarios.

(3) The gating-based fusion is also effective compared to concatenation, which provides personalized strategies in combining internal short-term and global long-term representations.

(4) The fourth ablation version removes the GAT-based aggregation in Eq. (\ref{eq.aggregation}) and the multi-hop neighbor-similarity based loss in Eq. (\ref{eq:neighbor_similarity_loss}) (only use the raw features of internal and external behaviors as inputs). The GAT-based aggregation and the multi-hop neighbor-similarity based loss enable more sufficient multi-domain user-item interactions, which are beneficial in capturing user variable and diverse preferences in practice.

\begin{table}[!hbtp]
\centering
\small
\begin{tabular}{l|c|c|c}
\toprule
models & period1 & period2 & period3  \\
\midrule
LSTTM (final) & \textbf{0.8395} & \textbf{0.8504} & \textbf{0.8502} \\
\midrule
  \quad -- temporal MAML & 0.8373 & 0.8428 & 0.8421 \\
  \quad -- user external behaviors  & 0.8385 & 0.8459 & 0.8459  \\
  \quad -- gating-based fusion  & 0.8378 & 0.8488 & 0.8489  \\
  \quad -- GAT aggregation \& $L_N$ & 0.8304 & 0.8410 & 0.8411 \\
\bottomrule
\end{tabular}
\caption{Ablation tests on NewsRec-21B.}
\label{tab:ablation_test}
\end{table}

\section{Conclusion and Future Work}

In this work, we propose an LSTTM for online recommendation, which captures user long-term and short-term preferences from internal/external behaviors. The temporal MAML enables fast adaptations to new topics in recommendation. The effectiveness of LSTTM is verified in offline and online real-world evaluations.

In the future, we will polish the temporal MAML to build a more robust adaptation, and transfer the idea of temporal MAML to other temporal tasks. We will also explore some enhanced combinations with other online learning and meta-learning methods.

\bibliographystyle{ACM-Reference-Format}
\bibliography{reference}


\begin{thebibliography}{41}


\ifx \showCODEN    \undefined \def \showCODEN     #1{\unskip}     \fi
\ifx \showDOI      \undefined \def \showDOI       #1{#1}\fi
\ifx \showISBNx    \undefined \def \showISBNx     #1{\unskip}     \fi
\ifx \showISBNxiii \undefined \def \showISBNxiii  #1{\unskip}     \fi
\ifx \showISSN     \undefined \def \showISSN      #1{\unskip}     \fi
\ifx \showLCCN     \undefined \def \showLCCN      #1{\unskip}     \fi
\ifx \shownote     \undefined \def \shownote      #1{#1}          \fi
\ifx \showarticletitle \undefined \def \showarticletitle #1{#1}   \fi
\ifx \showURL      \undefined \def \showURL       {\relax}        \fi
\providecommand\bibfield[2]{#2}
\providecommand\bibinfo[2]{#2}
\providecommand\natexlab[1]{#1}
\providecommand\showeprint[2][]{arXiv:#2}

\bibitem[\protect\citeauthoryear{An, Wu, Wu, Zhang, Liu, and Xie}{An
  et~al\mbox{.}}{2019}]%
        {an2019neural}
\bibfield{author}{\bibinfo{person}{Mingxiao An}, \bibinfo{person}{Fangzhao Wu},
  \bibinfo{person}{Chuhan Wu}, \bibinfo{person}{Kun Zhang},
  \bibinfo{person}{Zheng Liu}, {and} \bibinfo{person}{Xing Xie}.}
  \bibinfo{year}{2019}\natexlab{}.
\newblock \showarticletitle{Neural news recommendation with long-and short-term
  user representations}. In \bibinfo{booktitle}{\emph{Proceedings of ACL}}.
\newblock


\bibitem[\protect\citeauthoryear{Bharadhwaj}{Bharadhwaj}{2019}]%
        {bharadhwaj2019meta}
\bibfield{author}{\bibinfo{person}{Homanga Bharadhwaj}.}
  \bibinfo{year}{2019}\natexlab{}.
\newblock \showarticletitle{Meta-Learning for User Cold-Start Recommendation}.
  In \bibinfo{booktitle}{\emph{Proceedings of IJCNN}}.
\newblock


\bibitem[\protect\citeauthoryear{Dong, Yuan, Yao, Xu, and Zhu}{Dong
  et~al\mbox{.}}{2020}]%
        {dong2020mamo}
\bibfield{author}{\bibinfo{person}{Manqing Dong}, \bibinfo{person}{Feng Yuan},
  \bibinfo{person}{Lina Yao}, \bibinfo{person}{Xiwei Xu}, {and}
  \bibinfo{person}{Liming Zhu}.} \bibinfo{year}{2020}\natexlab{}.
\newblock \showarticletitle{MAMO: Memory-Augmented Meta-Optimization for
  Cold-start Recommendation}. In \bibinfo{booktitle}{\emph{Proceedings of
  KDD}}.
\newblock


\bibitem[\protect\citeauthoryear{Du, Wang, Yang, Zhou, and Tang}{Du
  et~al\mbox{.}}{2019}]%
        {du2019sequential}
\bibfield{author}{\bibinfo{person}{Zhengxiao Du}, \bibinfo{person}{Xiaowei
  Wang}, \bibinfo{person}{Hongxia Yang}, \bibinfo{person}{Jingren Zhou}, {and}
  \bibinfo{person}{Jie Tang}.} \bibinfo{year}{2019}\natexlab{}.
\newblock \showarticletitle{Sequential Scenario-Specific Meta Learner for
  Online Recommendation}. In \bibinfo{booktitle}{\emph{Proceedings of KDD}}.
\newblock


\bibitem[\protect\citeauthoryear{Duchi, Hazan, and Singer}{Duchi
  et~al\mbox{.}}{2011}]%
        {duchi2011adaptive}
\bibfield{author}{\bibinfo{person}{John Duchi}, \bibinfo{person}{Elad Hazan},
  {and} \bibinfo{person}{Yoram Singer}.} \bibinfo{year}{2011}\natexlab{}.
\newblock \showarticletitle{Adaptive subgradient methods for online learning
  and stochastic optimization.}
\newblock \bibinfo{journal}{\emph{JMLR}} (\bibinfo{year}{2011}).
\newblock


\bibitem[\protect\citeauthoryear{Finn, Abbeel, and Levine}{Finn
  et~al\mbox{.}}{2017}]%
        {finn2017model}
\bibfield{author}{\bibinfo{person}{Chelsea Finn}, \bibinfo{person}{Pieter
  Abbeel}, {and} \bibinfo{person}{Sergey Levine}.}
  \bibinfo{year}{2017}\natexlab{}.
\newblock \showarticletitle{Model-agnostic meta-learning for fast adaptation of
  deep networks}. In \bibinfo{booktitle}{\emph{Proceedings of ICML}}.
\newblock


\bibitem[\protect\citeauthoryear{Guo, Tang, Ye, Li, and He}{Guo
  et~al\mbox{.}}{2017}]%
        {guo2017deepfm}
\bibfield{author}{\bibinfo{person}{Huifeng Guo}, \bibinfo{person}{Ruiming
  Tang}, \bibinfo{person}{Yunming Ye}, \bibinfo{person}{Zhenguo Li}, {and}
  \bibinfo{person}{Xiuqiang He}.} \bibinfo{year}{2017}\natexlab{}.
\newblock \showarticletitle{DeepFM: a factorization-machine based neural
  network for CTR prediction}. In \bibinfo{booktitle}{\emph{Proceedings of
  IJCAI}}.
\newblock


\bibitem[\protect\citeauthoryear{He and Chua}{He and Chua}{2017}]%
        {he2017neural}
\bibfield{author}{\bibinfo{person}{Xiangnan He} {and} \bibinfo{person}{Tat-Seng
  Chua}.} \bibinfo{year}{2017}\natexlab{}.
\newblock \showarticletitle{Neural factorization machines for sparse predictive
  analytics}. In \bibinfo{booktitle}{\emph{Proceedings of SIGIR}}.
\newblock


\bibitem[\protect\citeauthoryear{Hu, Li, Shi, Yang, and Shao}{Hu
  et~al\mbox{.}}{2020}]%
        {hu2020graph}
\bibfield{author}{\bibinfo{person}{Linmei Hu}, \bibinfo{person}{Chen Li},
  \bibinfo{person}{Chuan Shi}, \bibinfo{person}{Cheng Yang}, {and}
  \bibinfo{person}{Chao Shao}.} \bibinfo{year}{2020}\natexlab{}.
\newblock \showarticletitle{Graph neural news recommendation with long-term and
  short-term interest modeling}.
\newblock \bibinfo{journal}{\emph{Information Processing \& Management}}
  (\bibinfo{year}{2020}).
\newblock


\bibitem[\protect\citeauthoryear{Lee, Im, Jang, Cho, and Chung}{Lee
  et~al\mbox{.}}{2019}]%
        {lee2019melu}
\bibfield{author}{\bibinfo{person}{Hoyeop Lee}, \bibinfo{person}{Jinbae Im},
  \bibinfo{person}{Seongwon Jang}, \bibinfo{person}{Hyunsouk Cho}, {and}
  \bibinfo{person}{Sehee Chung}.} \bibinfo{year}{2019}\natexlab{}.
\newblock \showarticletitle{MeLU: Meta-Learned User Preference Estimator for
  Cold-Start Recommendation}. In \bibinfo{booktitle}{\emph{Proceedings of
  KDD}}.
\newblock


\bibitem[\protect\citeauthoryear{Liu, Xie, Chen, Liu, Tu, Cui, Zhang, and
  Lin}{Liu et~al\mbox{.}}{2020b}]%
        {liu2020graph}
\bibfield{author}{\bibinfo{person}{Qi Liu}, \bibinfo{person}{Ruobing Xie},
  \bibinfo{person}{Lei Chen}, \bibinfo{person}{Shukai Liu}, \bibinfo{person}{Ke
  Tu}, \bibinfo{person}{Peng Cui}, \bibinfo{person}{Bo Zhang}, {and}
  \bibinfo{person}{Leyu Lin}.} \bibinfo{year}{2020}\natexlab{b}.
\newblock \showarticletitle{Graph Neural Network for Tag Ranking in
  Tag-enhanced Video Recommendation}. In \bibinfo{booktitle}{\emph{Proceedings
  of CIKM}}.
\newblock


\bibitem[\protect\citeauthoryear{Liu, Zeng, Mokhosi, and Zhang}{Liu
  et~al\mbox{.}}{2018}]%
        {liu2018stamp}
\bibfield{author}{\bibinfo{person}{Qiao Liu}, \bibinfo{person}{Yifu Zeng},
  \bibinfo{person}{Refuoe Mokhosi}, {and} \bibinfo{person}{Haibin Zhang}.}
  \bibinfo{year}{2018}\natexlab{}.
\newblock \showarticletitle{STAMP: short-term attention/memory priority model
  for session-based recommendation}. In \bibinfo{booktitle}{\emph{Proceedings
  of KDD}}.
\newblock


\bibitem[\protect\citeauthoryear{Liu, Chen, Sun, Xie, Gao, Ding, and Shen}{Liu
  et~al\mbox{.}}{2020a}]%
        {liu2020intent}
\bibfield{author}{\bibinfo{person}{Zhaoyang Liu}, \bibinfo{person}{Haokun
  Chen}, \bibinfo{person}{Fei Sun}, \bibinfo{person}{Xu Xie},
  \bibinfo{person}{Jinyang Gao}, \bibinfo{person}{Bolin Ding}, {and}
  \bibinfo{person}{Yanyan Shen}.} \bibinfo{year}{2020}\natexlab{a}.
\newblock \showarticletitle{Intent Preference Decoupling for User
  Representation on Online Recommender System}. In
  \bibinfo{booktitle}{\emph{Proceedings of IJCAI}}.
\newblock


\bibitem[\protect\citeauthoryear{Lu, Fang, and Shi}{Lu et~al\mbox{.}}{2020}]%
        {lu2020meta}
\bibfield{author}{\bibinfo{person}{Yuanfu Lu}, \bibinfo{person}{Yuan Fang},
  {and} \bibinfo{person}{Chuan Shi}.} \bibinfo{year}{2020}\natexlab{}.
\newblock \showarticletitle{Meta-learning on Heterogeneous Information Networks
  for Cold-start Recommendation}. In \bibinfo{booktitle}{\emph{Proceedings of
  KDD}}.
\newblock


\bibitem[\protect\citeauthoryear{Luo, Chen, Cheng, Dong, He, Feng, and Li}{Luo
  et~al\mbox{.}}{2020a}]%
        {luo2020metaselector}
\bibfield{author}{\bibinfo{person}{Mi Luo}, \bibinfo{person}{Fei Chen},
  \bibinfo{person}{Pengxiang Cheng}, \bibinfo{person}{Zhenhua Dong},
  \bibinfo{person}{Xiuqiang He}, \bibinfo{person}{Jiashi Feng}, {and}
  \bibinfo{person}{Zhenguo Li}.} \bibinfo{year}{2020}\natexlab{a}.
\newblock \showarticletitle{MetaSelector: Meta-Learning for Recommendation with
  User-Level Adaptive Model Selection}. In
  \bibinfo{booktitle}{\emph{Proceedings of WWW}}.
\newblock


\bibitem[\protect\citeauthoryear{Luo, Huang, Zhang, Wang, Baktashmotlagh, and
  Yang}{Luo et~al\mbox{.}}{2020b}]%
        {luo2020learning}
\bibfield{author}{\bibinfo{person}{Yadan Luo}, \bibinfo{person}{Zi Huang},
  \bibinfo{person}{Zheng Zhang}, \bibinfo{person}{Ziwei Wang},
  \bibinfo{person}{Mahsa Baktashmotlagh}, {and} \bibinfo{person}{Yang Yang}.}
  \bibinfo{year}{2020}\natexlab{b}.
\newblock \showarticletitle{Learning from the Past: Continual Meta-Learning
  with Bayesian Graph Neural Networks}. In
  \bibinfo{booktitle}{\emph{Proceedings of AAAI}}.
\newblock


\bibitem[\protect\citeauthoryear{Pan, Li, Ao, Tang, and He}{Pan
  et~al\mbox{.}}{2019}]%
        {pan2019warm}
\bibfield{author}{\bibinfo{person}{Feiyang Pan}, \bibinfo{person}{Shuokai Li},
  \bibinfo{person}{Xiang Ao}, \bibinfo{person}{Pingzhong Tang}, {and}
  \bibinfo{person}{Qing He}.} \bibinfo{year}{2019}\natexlab{}.
\newblock \showarticletitle{Warm up cold-start advertisements: Improving ctr
  predictions via learning to learn id embeddings}. In
  \bibinfo{booktitle}{\emph{Proceedings of SIGIR}}.
\newblock


\bibitem[\protect\citeauthoryear{Perozzi, Al-Rfou, and Skiena}{Perozzi
  et~al\mbox{.}}{2014}]%
        {perozzi2014deepwalk}
\bibfield{author}{\bibinfo{person}{Bryan Perozzi}, \bibinfo{person}{Rami
  Al-Rfou}, {and} \bibinfo{person}{Steven Skiena}.}
  \bibinfo{year}{2014}\natexlab{}.
\newblock \showarticletitle{Deepwalk: Online learning of social
  representations}. In \bibinfo{booktitle}{\emph{Proceedings of KDD}}.
\newblock


\bibitem[\protect\citeauthoryear{Rendle}{Rendle}{2010}]%
        {rendle2010factorization}
\bibfield{author}{\bibinfo{person}{Steffen Rendle}.}
  \bibinfo{year}{2010}\natexlab{}.
\newblock \showarticletitle{Factorization machines}. In
  \bibinfo{booktitle}{\emph{Proceedings of ICDM}}.
\newblock


\bibitem[\protect\citeauthoryear{Song, Shi, Xiao, Duan, Xu, Zhang, and
  Tang}{Song et~al\mbox{.}}{2019}]%
        {song2019autoint}
\bibfield{author}{\bibinfo{person}{Weiping Song}, \bibinfo{person}{Chence Shi},
  \bibinfo{person}{Zhiping Xiao}, \bibinfo{person}{Zhijian Duan},
  \bibinfo{person}{Yewen Xu}, \bibinfo{person}{Ming Zhang}, {and}
  \bibinfo{person}{Jian Tang}.} \bibinfo{year}{2019}\natexlab{}.
\newblock \showarticletitle{Autoint: Automatic feature interaction learning via
  self-attentive neural networks}. In \bibinfo{booktitle}{\emph{Proceedings of
  CIKM}}.
\newblock


\bibitem[\protect\citeauthoryear{Sun, Liu, Wu, Pei, Lin, Ou, and Jiang}{Sun
  et~al\mbox{.}}{2019}]%
        {sun2019bert4rec}
\bibfield{author}{\bibinfo{person}{Fei Sun}, \bibinfo{person}{Jun Liu},
  \bibinfo{person}{Jian Wu}, \bibinfo{person}{Changhua Pei},
  \bibinfo{person}{Xiao Lin}, \bibinfo{person}{Wenwu Ou}, {and}
  \bibinfo{person}{Peng Jiang}.} \bibinfo{year}{2019}\natexlab{}.
\newblock \showarticletitle{BERT4Rec: Sequential Recommendation with
  Bidirectional Encoder Representations from Transformer}. In
  \bibinfo{booktitle}{\emph{Proceedings of CIKM}}.
\newblock


\bibitem[\protect\citeauthoryear{Vartak, Thiagarajan, Miranda, Bratman, and
  Larochelle}{Vartak et~al\mbox{.}}{2017}]%
        {vartak2017meta}
\bibfield{author}{\bibinfo{person}{Manasi Vartak}, \bibinfo{person}{Arvind
  Thiagarajan}, \bibinfo{person}{Conrado Miranda}, \bibinfo{person}{Jeshua
  Bratman}, {and} \bibinfo{person}{Hugo Larochelle}.}
  \bibinfo{year}{2017}\natexlab{}.
\newblock \showarticletitle{A meta-learning perspective on cold-start
  recommendations for items}. In \bibinfo{booktitle}{\emph{Proceedings of
  NIPS}}.
\newblock


\bibitem[\protect\citeauthoryear{Veli{\v{c}}kovi{\'c}, Cucurull, Casanova,
  Romero, Lio, and Bengio}{Veli{\v{c}}kovi{\'c} et~al\mbox{.}}{2018}]%
        {velivckovic2018graph}
\bibfield{author}{\bibinfo{person}{Petar Veli{\v{c}}kovi{\'c}},
  \bibinfo{person}{Guillem Cucurull}, \bibinfo{person}{Arantxa Casanova},
  \bibinfo{person}{Adriana Romero}, \bibinfo{person}{Pietro Lio}, {and}
  \bibinfo{person}{Yoshua Bengio}.} \bibinfo{year}{2018}\natexlab{}.
\newblock \showarticletitle{Graph attention networks}. In
  \bibinfo{booktitle}{\emph{Proceedings of ICLR}}.
\newblock


\bibitem[\protect\citeauthoryear{Vilalta and Drissi}{Vilalta and
  Drissi}{2002}]%
        {vilalta2002perspective}
\bibfield{author}{\bibinfo{person}{Ricardo Vilalta} {and}
  \bibinfo{person}{Youssef Drissi}.} \bibinfo{year}{2002}\natexlab{}.
\newblock \showarticletitle{A perspective view and survey of meta-learning}.
\newblock \bibinfo{journal}{\emph{Artificial intelligence review}}
  (\bibinfo{year}{2002}).
\newblock


\bibitem[\protect\citeauthoryear{Wang, Wei, Cong, Li, Mao, and Qiu}{Wang
  et~al\mbox{.}}{2020}]%
        {wang2020global}
\bibfield{author}{\bibinfo{person}{Ziyang Wang}, \bibinfo{person}{Wei Wei},
  \bibinfo{person}{Gao Cong}, \bibinfo{person}{Xiao-Li Li},
  \bibinfo{person}{Xian-Ling Mao}, {and} \bibinfo{person}{Minghui Qiu}.}
  \bibinfo{year}{2020}\natexlab{}.
\newblock \showarticletitle{Global context enhanced graph neural networks for
  session-based recommendation}. In \bibinfo{booktitle}{\emph{Proceedings of
  SIGIR}}.
\newblock


\bibitem[\protect\citeauthoryear{Wu, Tang, Zhu, Wang, Xie, and Tan}{Wu
  et~al\mbox{.}}{2019}]%
        {wu2019session}
\bibfield{author}{\bibinfo{person}{Shu Wu}, \bibinfo{person}{Yuyuan Tang},
  \bibinfo{person}{Yanqiao Zhu}, \bibinfo{person}{Liang Wang},
  \bibinfo{person}{Xing Xie}, {and} \bibinfo{person}{Tieniu Tan}.}
  \bibinfo{year}{2019}\natexlab{}.
\newblock \showarticletitle{Session-based Recommendation with Graph Neural
  Networks}. In \bibinfo{booktitle}{\emph{Proceedings of AAAI}}.
\newblock


\bibitem[\protect\citeauthoryear{Xiang, Yuan, Zhao, Chen, Zhang, Yang, and
  Sun}{Xiang et~al\mbox{.}}{2010}]%
        {xiang2010temporal}
\bibfield{author}{\bibinfo{person}{Liang Xiang}, \bibinfo{person}{Quan Yuan},
  \bibinfo{person}{Shiwan Zhao}, \bibinfo{person}{Li Chen},
  \bibinfo{person}{Xiatian Zhang}, \bibinfo{person}{Qing Yang}, {and}
  \bibinfo{person}{Jimeng Sun}.} \bibinfo{year}{2010}\natexlab{}.
\newblock \showarticletitle{Temporal recommendation on graphs via long-and
  short-term preference fusion}. In \bibinfo{booktitle}{\emph{Proceedings of
  KDD}}.
\newblock


\bibitem[\protect\citeauthoryear{Xiao, Xie, Yao, Liu, Sun, Zhang, and Lin}{Xiao
  et~al\mbox{.}}{2021}]%
        {xiao2021uprec}
\bibfield{author}{\bibinfo{person}{Chaojun Xiao}, \bibinfo{person}{Ruobing
  Xie}, \bibinfo{person}{Yuan Yao}, \bibinfo{person}{Zhiyuan Liu},
  \bibinfo{person}{Maosong Sun}, \bibinfo{person}{Xu Zhang}, {and}
  \bibinfo{person}{Leyu Lin}.} \bibinfo{year}{2021}\natexlab{}.
\newblock \showarticletitle{UPRec: User-Aware Pre-training for Recommender
  Systems}.
\newblock \bibinfo{journal}{\emph{arXiv preprint arXiv:2102.10989}}
  (\bibinfo{year}{2021}).
\newblock


\bibitem[\protect\citeauthoryear{Xie, Ling, Wang, Wang, Xia, and Lin}{Xie
  et~al\mbox{.}}{2020a}]%
        {xie2020deep}
\bibfield{author}{\bibinfo{person}{Ruobing Xie}, \bibinfo{person}{Cheng Ling},
  \bibinfo{person}{Yalong Wang}, \bibinfo{person}{Rui Wang},
  \bibinfo{person}{Feng Xia}, {and} \bibinfo{person}{Leyu Lin}.}
  \bibinfo{year}{2020}\natexlab{a}.
\newblock \showarticletitle{Deep Feedback Network for Recommendation}. In
  \bibinfo{booktitle}{\emph{Proceedings of IJCAI}}.
\newblock


\bibitem[\protect\citeauthoryear{Xie, Liu, Liu, Zhang, Cui, Zhang, and Lin}{Xie
  et~al\mbox{.}}{2021a}]%
        {xie2021improving}
\bibfield{author}{\bibinfo{person}{Ruobing Xie}, \bibinfo{person}{Qi Liu},
  \bibinfo{person}{Shukai Liu}, \bibinfo{person}{Ziwei Zhang},
  \bibinfo{person}{Peng Cui}, \bibinfo{person}{Bo Zhang}, {and}
  \bibinfo{person}{Leyu Lin}.} \bibinfo{year}{2021}\natexlab{a}.
\newblock \showarticletitle{Improving Accuracy and Diversity in Matching of
  Recommendation with Diversified Preference Network}.
\newblock \bibinfo{journal}{\emph{IEEE Transactions on Big Data}}
  (\bibinfo{year}{2021}).
\newblock


\bibitem[\protect\citeauthoryear{Xie, Qiu, Rao, Liu, Zhang, and Lin}{Xie
  et~al\mbox{.}}{2020b}]%
        {xie2020internal}
\bibfield{author}{\bibinfo{person}{Ruobing Xie}, \bibinfo{person}{Zhijie Qiu},
  \bibinfo{person}{Jun Rao}, \bibinfo{person}{Yi Liu}, \bibinfo{person}{Bo
  Zhang}, {and} \bibinfo{person}{Leyu Lin}.} \bibinfo{year}{2020}\natexlab{b}.
\newblock \showarticletitle{Internal and Contextual Attention Network for
  Cold-start Multi-channel Matching in Recommendation}. In
  \bibinfo{booktitle}{\emph{Proceedings of IJCAI}}.
\newblock


\bibitem[\protect\citeauthoryear{Xie, Zhang, Wang, Xia, and Lin}{Xie
  et~al\mbox{.}}{2021b}]%
        {xie2021hierarchical}
\bibfield{author}{\bibinfo{person}{Ruobing Xie}, \bibinfo{person}{Shaoliang
  Zhang}, \bibinfo{person}{Rui Wang}, \bibinfo{person}{Feng Xia}, {and}
  \bibinfo{person}{Leyu Lin}.} \bibinfo{year}{2021}\natexlab{b}.
\newblock \showarticletitle{Hierarchical Reinforcement Learning for Integrated
  Recommendation}. In \bibinfo{booktitle}{\emph{Proceedings of AAAI}}.
\newblock


\bibitem[\protect\citeauthoryear{Yu, Lian, Mahmoody, Liu, and Xie}{Yu
  et~al\mbox{.}}{2019}]%
        {yu2019adaptive}
\bibfield{author}{\bibinfo{person}{Zeping Yu}, \bibinfo{person}{Jianxun Lian},
  \bibinfo{person}{Ahmad Mahmoody}, \bibinfo{person}{Gongshen Liu}, {and}
  \bibinfo{person}{Xing Xie}.} \bibinfo{year}{2019}\natexlab{}.
\newblock \showarticletitle{Adaptive User Modeling with Long and Short-Term
  Preferences for Personalized Recommendation.}. In
  \bibinfo{booktitle}{\emph{Proceedings of IJCAI}}.
\newblock


\bibitem[\protect\citeauthoryear{Zeng, Xiao, Yao, Xie, Liu, Lin, Lin, and
  Sun}{Zeng et~al\mbox{.}}{2021}]%
        {zeng2021knowledge}
\bibfield{author}{\bibinfo{person}{Zheni Zeng}, \bibinfo{person}{Chaojun Xiao},
  \bibinfo{person}{Yuan Yao}, \bibinfo{person}{Ruobing Xie},
  \bibinfo{person}{Zhiyuan Liu}, \bibinfo{person}{Fen Lin},
  \bibinfo{person}{Leyu Lin}, {and} \bibinfo{person}{Maosong Sun}.}
  \bibinfo{year}{2021}\natexlab{}.
\newblock \showarticletitle{Knowledge transfer via pre-training for
  recommendation: A review and prospect}.
\newblock \bibinfo{journal}{\emph{Frontiers in big Data}}
  (\bibinfo{year}{2021}).
\newblock


\bibitem[\protect\citeauthoryear{Zhang, Feng, Wang, He, Wang, Li, and
  Zhang}{Zhang et~al\mbox{.}}{2020}]%
        {zhang2020retrain}
\bibfield{author}{\bibinfo{person}{Yang Zhang}, \bibinfo{person}{Fuli Feng},
  \bibinfo{person}{Chenxu Wang}, \bibinfo{person}{Xiangnan He},
  \bibinfo{person}{Meng Wang}, \bibinfo{person}{Yan Li}, {and}
  \bibinfo{person}{Yongdong Zhang}.} \bibinfo{year}{2020}\natexlab{}.
\newblock \showarticletitle{How to Retrain Recommender System? A Sequential
  Meta-Learning Method}. In \bibinfo{booktitle}{\emph{Proceedings of SIGIR}}.
\newblock


\bibitem[\protect\citeauthoryear{Zhou, Mou, Fan, Pi, Bian, Zhou, Zhu, and
  Gai}{Zhou et~al\mbox{.}}{2019}]%
        {zhou2019deep}
\bibfield{author}{\bibinfo{person}{Guorui Zhou}, \bibinfo{person}{Na Mou},
  \bibinfo{person}{Ying Fan}, \bibinfo{person}{Qi Pi}, \bibinfo{person}{Weijie
  Bian}, \bibinfo{person}{Chang Zhou}, \bibinfo{person}{Xiaoqiang Zhu}, {and}
  \bibinfo{person}{Kun Gai}.} \bibinfo{year}{2019}\natexlab{}.
\newblock \showarticletitle{Deep interest evolution network for click-through
  rate prediction}. In \bibinfo{booktitle}{\emph{Proceedings of AAAI}}.
\newblock


\bibitem[\protect\citeauthoryear{Zhou, Zhu, Song, Fan, Zhu, Ma, Yan, Jin, Li,
  and Gai}{Zhou et~al\mbox{.}}{2018}]%
        {zhou2018deep}
\bibfield{author}{\bibinfo{person}{Guorui Zhou}, \bibinfo{person}{Xiaoqiang
  Zhu}, \bibinfo{person}{Chenru Song}, \bibinfo{person}{Ying Fan},
  \bibinfo{person}{Han Zhu}, \bibinfo{person}{Xiao Ma},
  \bibinfo{person}{Yanghui Yan}, \bibinfo{person}{Junqi Jin},
  \bibinfo{person}{Han Li}, {and} \bibinfo{person}{Kun Gai}.}
  \bibinfo{year}{2018}\natexlab{}.
\newblock \showarticletitle{Deep interest network for click-through rate
  prediction}. In \bibinfo{booktitle}{\emph{Proceedings of KDD}}.
\newblock


\bibitem[\protect\citeauthoryear{Zhu, Ge, Zhuang, Xie, Xi, Zhang, Lin, and
  He}{Zhu et~al\mbox{.}}{2021a}]%
        {zhu2021transfer}
\bibfield{author}{\bibinfo{person}{Yongchun Zhu}, \bibinfo{person}{kaikai Ge},
  \bibinfo{person}{Fuzhen Zhuang}, \bibinfo{person}{Ruobing Xie},
  \bibinfo{person}{Dongbo Xi}, \bibinfo{person}{Xu Zhang},
  \bibinfo{person}{Leyu Lin}, {and} \bibinfo{person}{Qing He}.}
  \bibinfo{year}{2021}\natexlab{a}.
\newblock \showarticletitle{Transfer-Meta Framework for Cross-domain
  Recommendation to Cold-Start Users}. In \bibinfo{booktitle}{\emph{Proceedings
  of SIGIR}}.
\newblock


\bibitem[\protect\citeauthoryear{Zhu, Liu, Xie, Zhuang, Hao, Ge, Zhang, Lin,
  and Cao}{Zhu et~al\mbox{.}}{2021b}]%
        {zhu2021learn}
\bibfield{author}{\bibinfo{person}{Yongchun Zhu}, \bibinfo{person}{Yudan Liu},
  \bibinfo{person}{Ruobing Xie}, \bibinfo{person}{Fuzhen Zhuang},
  \bibinfo{person}{Xiaobo Hao}, \bibinfo{person}{Kaikai Ge},
  \bibinfo{person}{Xu Zhang}, \bibinfo{person}{Leyu Lin}, {and}
  \bibinfo{person}{Juan Cao}.} \bibinfo{year}{2021}\natexlab{b}.
\newblock \showarticletitle{Learn to Expand Audience via Meta Hybrid Experts
  and Critics for Recommendation and Advertising}. In
  \bibinfo{booktitle}{\emph{Proceedings of KDD}}.
\newblock


\bibitem[\protect\citeauthoryear{Zhu, Xie, Zhuang, Ge, Sun, Zhang, Lin, and
  Cao}{Zhu et~al\mbox{.}}{2021c}]%
        {zhu2021learning}
\bibfield{author}{\bibinfo{person}{Yongchun Zhu}, \bibinfo{person}{Ruobing
  Xie}, \bibinfo{person}{Fuzhen Zhuang}, \bibinfo{person}{Kaikai Ge},
  \bibinfo{person}{Ying Sun}, \bibinfo{person}{Xu Zhang}, \bibinfo{person}{Leyu
  Lin}, {and} \bibinfo{person}{Juan Cao}.} \bibinfo{year}{2021}\natexlab{c}.
\newblock \showarticletitle{Learning to Warm Up Cold Item Embeddings for
  Cold-start Recommendation with Meta Scaling and Shifting Networks}. In
  \bibinfo{booktitle}{\emph{Proceedings of SIGIR}}.
\newblock


\bibitem[\protect\citeauthoryear{Z{\"u}gner and G{\"u}nnemann}{Z{\"u}gner and
  G{\"u}nnemann}{2019}]%
        {zugner2019adversarial}
\bibfield{author}{\bibinfo{person}{Daniel Z{\"u}gner} {and}
  \bibinfo{person}{Stephan G{\"u}nnemann}.} \bibinfo{year}{2019}\natexlab{}.
\newblock \showarticletitle{Adversarial attacks on graph neural networks via
  meta learning}. In \bibinfo{booktitle}{\emph{Proceedings of ICLR}}.
\newblock


\end{thebibliography}

\end{document}